\documentclass[12pt,preprint]{aastex}

\shorttitle{\ion{N}{7} Hyperfine Line Observations}
\shortauthors{Bregman and Irwin}

\begin{document}


\title{The Search for Million Degree Gas Through The \ion{N}{7} Hyperfine Line}

\author{Joel N. Bregman and Jimmy A. Irwin}
\affil{Department of Astronomy, University of Michigan, Ann Arbor, MI 48109}
\email{jbregman@umich.edu, jairwin@umich.edu}

\begin{abstract}

Gas in the million degree range occurs in a variety of astronomical
environments, and it may be the main component
of the elusive missing baryons at low redshift.  The \ion{N}{7} ion is found in
this material and it has a hyperfine spin-flip transition with a rest
frequency of 53.042 GHz, which can be observed for z $>$ 0.1, when it is
shifted into a suitably transparent radio band.  We used the 42-48 GHz
spectrometer on the {\it Green Bank Telescope\/} to search for both emission
and absorption from this \ion{N}{7} transmission.  For absorption studies,
3C273, 3C 279, 3C 345, and 4C+39.25 were observed but no feature were seen 
above the 5$\sigma$ level.  For emission line studies, we observed Abell 1835,
Abell 2390 and the star-forming galaxy PKS 1345+12, but no features were seen exceeding 
5$\sigma$.  We examine whether the strongest emission feature, in Abell 2390 (3.7$\sigma$),
and the strongest absorption feature, toward 4C+39.25 (3.8$\sigma$), might be
expected from theoretical models.  The emission feature would require 
$\sim$10$^{{\rm 10}}$ M$_{\odot}$ of 10$^{{\rm 6}}$ K gas, which is inconsistent
with X-ray limits for the \ion{O}{7} K$\alpha$ line, so it is unlikely to be real.
The \ion{N}{7} absorption feature requires a \ion{N}{7} column of 
6$\times$10$^{{\rm 16}}$ cm$^{{\rm -}{2}}$, higher than model predictions by at
least an order of magnitude, which makes it inconsistent with model expectations.
The individual observations were less than 1 hr in length, so for lengthy 
observations, we show that \ion{N}{7} absorption line observations can
begin to be useful in in the search for hot intergalactic gas.

\end{abstract}


\section{Introduction}

The study of hot gas at temperature of 10$^{{\rm 6}}$ K or greater has been the
province of X-ray astronomy since the resonance transitions of the most
common ionic states occur in the 0.1-8 keV range.  Hot gas is found to
be very common in a wide range of astronomical objects, such as galaxy
clusters, early-type galaxies, and starburst galaxies, where it is seen in
emission.  In addition, most of the baryons in the local universe are
predicted to have temperatures of 10$^{{\rm 5}}$ - 10$^{{\rm 7}}$ K, filling the volumes of
modest overdensity cosmological filaments, sometimes referred to as the
``Cosmic Web'' (this is also the warm-hot intergalactic medium, or
WHIM).  This gas is too dilute to be a significant emitter, but it is
possible for it to produce absorption, which has been searched for at 
X-ray wavelengths, but not convincingly detected \citep{rasm06}.

The radio band provides another approach to detecting hot gas through
the hyperfine transitions of ions with magnetic moments, a possibility first
discussed by \citet{suny84} in the context of supernova remnants and
galaxy clusters.  An advantage to using this approach is that the velocity
resolution is far superior relative to current X-ray techniques, permitting one to easily
resolve the hyperfine line and obtain a precise line center.  The primary
disadvantage is that most of the lines are weak and sometimes at
frequencies where it is difficult to make observations.  

This transition is a hyperfine spin-flip transition, the same that is
responsible for the 21 cm line of neutral hydrogen.  It requires that the
magnetic moment of an electron interact with the magnetic moment of the
nucleus.  The magnetic moment is greatest for the electron if there is a
single valence electron, so hydrogenic or lithium-like ions are preferred,
along with nuclei that have an odd number of protons (or an odd mass
number) so that not all the magnetic moments are cancelled.  The first
criteria is easily met, but the second one is more problematic because
nature prefers to build atoms with alpha particles, so nearly all of the
abundant elements have an even number nuclear charge or an even atomic
mass.  The two most common elements with odd nuclear charge are
hydrogen and nitrogen, the latter being abundant only due to the presence
of the CNO cycle in stars.  Therefore, it is hydrogenic-like nitrogen
(\ion{N}{7}) that has the most favorable hyperfine transition, after the HI 21
cm line \citep{godd03,suny06}.  Unfortunately, the line from $^{{\rm 14}}$N occurs at 53.042
GHz \citep{shab95}, a frequency regime where there is very significant attenuation due
to oxygen resonance lines in the Earth's atmosphere, making low redshift
observations impossible.

Because of this opacity problem, as well as the lack of receivers, most of
the attention of observers was focused on the Li-like Fe isotope line, $^{{\rm 57}}$Fe
XXIV, which is present in gas near 10$^{{\rm 7}}$ K and has a transition at 98 GHz
\citep{dcru98}, a clear mm window for which there are good receivers. 
This line was  searched for in emission, but not detected (Liang et al.
1997), the limits not being very restrictive in regard to model predictions
\citep{godd03}.

Here we reconsider the emission and absorption possibilities for $^{{\rm 14}}$N,
which is two orders of magnitude more common than $^{{\rm 57}}$Fe.  Our
motivation is that the {\it Green Bank Telescope \/}(GBT) has an excellent
receiver in the atmospherically transparent 42-48 GHz range, with a
spectrometer that can sample this entire range ($\bigtriangleup$ z = 0.16) in only
four separate settings and with a velocity resolution of 2.7 km s$^{{\rm -1}}$, which
could easily resolve emission or absorption from this ion at its Doppler
width (FWHM = 80 km s$^{{\rm -1}}$ at 10$^{{\rm 6}}$ K).  To use this receiver, we are forced to work in
the redshift range 0.11 - 0.26.

This $^{{\rm 14}}$N hyperfine line has atomic properties that make it competitive
with the HI hyperfine line for both emission and absorption.  It has a
critical density of 0.15 cm$^{{\rm -3}}$, and above this value, the emissivity is $\epsilon$ $\propto$
n$_{{\rm i}}$A$_{{\rm 21}}$, where n$_{{\rm i}}$ is the density of 
an element in a particular ionization state
and A$_{{\rm 21}}$ is the Einstein-A value.  
For hydrogen, A$_{{\rm 21}}$= 2.87$\times$10$^{{\rm -}{15}}$ sec$^{{\rm -}{1}}$,
while for the \ion{N}{7} line, A$_{{\rm 21}}$ = 1.99$\times$10$^{{\rm -}{10}}$ sec$^{{\rm -}{1}}$ 
\citep{suny06}, so even though the nitrogen
abundance is 9.1$\times$10$^{{\rm -}{5}}$ relative to hydrogen (at Solar abundance), the
difference in the A values compensates so that the volume emissivity is
the same for \ion{N}{7} line and the 21 cm line when nitrogen is 1/6 of its
Solar abundance (comparing equal masses of 10$^{{\rm 6}}$ K gas and HI gas).  The
\ion{N}{7} emission line will be detectable in this redshift range if 10$^{{\rm 10}}$-10$^{{\rm 11}}$
M$_{\odot}$ of million degree gas is present, which might occur in a cluster of
galaxies or an ultraluminous infrared galaxy (ULIRG).  The emission line component of our GBT program
was to search for such emission from a few different objects.

In absorption, the density of the WHIM is predicted to be 10$^{{\rm -4 }}$- 10$^{{\rm -5}}$ cm$^{{\rm -}{3}}$
\citep{hell98}, well below the critical density, so all atoms are in their
ground state.  Thus, unlike the HI 21 cm transition, there is no stimulated
emission to reduce the optical depth.  The \ion{N}{7} line becomes detectable in
absorption for a column density of 10$^{{\rm 20}}$ -10$^{{\rm 21}}$ cm$^{{\rm -2}}$ 
of million degree gas (1/10 Solar metallicities), which would occur in a filament 1 Mpc in
length and of total particle density 10$^{{\rm -4}}$ cm$^{{\rm -}{3}}$.  Such column densities can
occur, according to simulations \citep{hell98}, so we have searched for
absorption features against strong continuum sources in the 42-48 GHz
range.

\section{Data Processing}

The observations were taken using the 100 m NRAO Green Bank Telescope (GBT) during the
two nights of 2006 February 19 and February 21 in the 42--48 GHz frequency
range using the Q-band receiver. The Q-band receiver is divided into two
frequency ranges with separate feed and amplifier sets. The GBT spectrometer
was configured in its four-intermediate frequncy (IF) 800 MHz width
three-level mode, which allows observing two 800 MHz frequency bands in
two polarizations at once. Double beam switching was employed such that the
source appeared in each of the dual beams alternately for a period of
1 minute.
The data were analyzed using GBTIDL \citep{garw05} v1.2.1. Bad scans
were removed, and the remaining scans were accumulated and averaged.
The total useful observing time for each object/band number is listed in
Table 1. 

Most of the observing time was used to search for absorption lines
against strong continuum sources, but the baselines cannot be fit with
simple functions, which introduces challenges.  Presumably, the structure
in the baseline is due to standing waves in the feedhorn and receiver, but
this is not well-documented yet.  To flatten the baseline structure, we
tried a variety of procedures, the most successful being a two-step
procedure.  In the first step, we smooth the spectrum, using adjacent
averaging over 200 channels, or about 520 km s$^{{\rm -1}}$ at 45 GHz (Figure 1). 
The original spectrum, in which T$_{{\rm a}}$ varies by about 10\% of a 0.8 GHz
band, is divided by the smoothed spectrum, and this reduces the variation
by more than an order of magnitude, to less than 1\% (Figure 2). 
However, most of these variations are systematic in nature and are nearly
identical for two different continuum sources observed over similar
frequency ranges.  These systematics were largely removed by subtracting
the fluctuations about unity in one spectrum from the other spectrum, leading to a
final rms per bin (an average of 10 channels, or 26 km s$^{{\rm -}{1)}}$ 
that was often less than 0.1\% of the normalized flux density (Figure 2).

We tested this procedure by adding spectral lines to a source (3C345) to
see if they were correctly recovered.  The tests worked well and we
successfully recovered absorption lines with central optical depths of
0.002 and widths of 90 km s$^{{\rm -1}}$ and 220 km s$^{{\rm -1}}$ (FWHM), which are about
the line widths anticipated (Figure 3).  Lines as wide as the averaging
length will not be recovered reliably, but we should easily be able to detect lines
of width 300 km s$^{{\rm -1}}$ or less.

One aspect of this procedure is that for the final step, one would ideally
want to have a strong featureless source, but in general, such a source
was either not available or not used.  Instead, for a calibration source, we
used either 3C 279 or 3C 273, or when both were observed in the same
band, the average of the two.  When using these sources for the final
calibration, a problem that can arise if one of the sources has an intrinsic
absorption line.  This will appear as an emission line in the final spectrum,
but this emission line will be present in all final spectra, and as there are
either three or four strong continuum sources observed in every band, one
can inspect the spectra for this effect, and we have done so.  
This two-step data processing method was used for 3C273, 3C279, 4C39.25,
3C345, and PKS 1345+12, which has a significantly weaker continuum
than the others.

For sources with a weaker continuum or without a detectable continuum,
we found that subtracting the smoothed continuum from the unsmoothed
data worked well.  This approach was applied to Abell 1835, Abell 2390,
and PKS 0405-12 (at continuum levels of 0.3 K), although for this source,
using the ratio method led to similar results.  For Abell 1835, there was a
noticeable slow baseline structure after subtracting the smoothed
continuum, so we also fit and removed a third order polynomial fit to the
baseline.  When observing Abell 2390, PKS 0405-12, and PKS 1345+12,
we used overlapping spectral bands to check that if interesting features
were observed, that the same structures were detected in both
observations.

\section{Results}

Most of the observations involved searching for absorption features against
the radio continua, but for two clusters of galaxies and one star-forming
galaxy, we searched for \ion{N}{7} emission.

PKS0405-12: Although the continuum is much lower than most other
targets, this was chosen because it has a known CIV and Ly$\alpha$ absorption
system at z = 0.167 \citep{spin93}, which would place a \ion{N}{7} absorption
line at 45.45 GHz.  Two sets of observations were obtained with the central
frequency shifted by 0.1 GHz and the central frequency was different than
that of the various strong continuum objects.  Possibly for this reason, we
were unable to use the second step of our continuum flattening
procedure, so the resulting rms is relatively higher than for the strong
continuum sources (Figure 4).  There is no evidence for any absorption at
the frequency of the \ion{N}{7} transition.  The upper limit is not in conflict
with the detection of the UV absorption lines, which are sensitive to much
lower column densities.

PKS1345+12: This strong IRAS source is similar to Arp 220 in that it has
a double nucleus and a great deal of star formation \citep{xian02}.  It is
also mentioned as a Seyfert 2 galaxy and an IR quasar.  We hoped to
search for emission due to the hot gas from star formation, or for
absorption by hot gas against the radio continuum.  Two adjacent bands
were observed, which included the redshift of the object, but neither
emission nor absorption were detected (Figure 5).  An object with a
continuum makes it more challenging to detect emission, so the rms is
larger than for objects without a significant continuum.

Abell 1835: This cluster of galaxies has strongly centrally concentrated 
X-ray emission with a cooling time much less than a Hubble time, so it was a
classic cooling flow cluster until {\it XMM\/} observations failed to find the
expected strong OVII emission \citep{allen01b, pete03}.  The {\it XMM\/} observations
place a limit on the cooling rate of $\sim$50 M$_{\odot}$ yr$^{-1}$.  There is no significant
radio continuum and there is no detectable emission from the \ion{N}{7} ion
(Figure 6).

Abell 2390: This is another cluster with strong X-ray emission, similar to
Abell 1835, but not as luminous and with a lower cooling rate as inferred
from pre-{\it XMM\/} data \citep{allen01a, allen01b, pete03}; H$_2$ was 
detected by \citet{edge02}.  
Two observations were obtained where the central
frequency was shifted by 0.3 GHz, but both scans included the redshift
region of Abell 2390.  The common regions were combined and the rms
was quite low in the resulting spectrum.  No features ar detected at or
above the 5$\sigma$ level, which we require for a confident detection.
The strongest emission feature is at the 3.7$\sigma$ level and near 
the redshift of the galaxy cluster (line flux
of 0.51$\pm$0.14 K km s$^{{\rm -1}}$, FWHM = 66$\pm$20 km s$^{{\rm -1}}$; 
Figure 7).  Statistically, a positive feature of
this significance occurs 0.01\% of the time.  The feature is present in both
of the individual scans and at about the same strength (Figure 8).  A
feature of this strength would have been below the detection threshold in
Abell 1835 or PKS 1345+12 due to the higher rms in the spectra.  We
hope to obtain addition observations to determine the reality of this
feature.

The Strong Continuum Sources: The two strong continuum sources 3C
279 and 3C 273 provide the highest S/N continua against which to search
for \ion{N}{7} absorption lines.  There are no known strong absorption lines in
either source over the available redshift range 0.11 $<$ z $<$ 0.26, although
the emission line redshift for 3C 273 lies in the observing range, at 45.8
GHz.  We did not observe 3C273 at frequencies below 45 GHz, so for
the 42-45 GHz region, we differenced 3C 279 with 3C 345 and 4C
+39.25 to search for absorption features.  Since we are differencing
spectra, should one have an absorption line, we would see that as an
emission line.  In the differences of 3C 279 and 3C 273 over the 45-48
GHz band, there were no statistically significant absorption or emission
features, with impressively low rms values (5$\sigma$ upper limits of 0.085 -
0.25 km s$^{{\rm -1}}$ for 100 km s$^{{\rm -1}}$ wide lines; Table 1 and Figure 9-12).  Since no
features were seen in this spectral region, we averaged the two spectra
and used that average to remove the small variations in the spectra of
3C345 and 4C+39.45 in the 45-48 GHz range.

In the frequency range 45-48 GHz, there are no significant absorption
feature in 3C 345 or 4C+39.25 above the 5$\sigma $ level, so we do not claim
detections due to absorption in any or our targets.
The strongest feature is found in absorption in the spectrum of 
4C+39.25, and it occurs in the middle of the 45.0-45.8 GHz band, which has the best response and
lowest rms of any band.  When we fit for the continuum level, central
frequency (45.38$\pm$0.04 GHz), line width (FWHM = 99$\pm$23 km s$^{{\rm -1}}$),
and equivalent width (0.19$\pm$0.05 km s$^{{\rm -1}}$), the feature has a significance of a 3.8$\sigma$. 
Had we fixed some fitting parameters, such as the continuum level, or not
fit for certain parameters, such as the line width, the significance would be
greater (4.9$\sigma$).  An absorption feature of this strength (3.8$\sigma$) would occur
by chance once in about 10$^{{\rm 4}}$ trials, provided that the noise has a normal
distribution.  There are about 30-40 independent
regions in a single 0.8 GHz band, so for all the bands of all the objects,
there were about 10$^{{\rm 3}}$ independent regions in which an absorption feature
could have occurred.  This feature is rarer than would be expected by
chance, but to be cautious, we intepret it as a non-detection until confirmed.

In the frequency range 42-45 GHz, there are no significant features in any
object (Figures 13-16).  There are a few features in the 2.5-3$\sigma$ range, but
that is to be expected for the amount of data obtained.

\section{Discussion and Final Comments}

Although we do not detect any emission or absorption features above the 
5$\sigma$ level, we can discuss whether the two strongest features are
likely to be confirmed.  For this exercise, we compare these features to
theoretical predictions.

We can convert the putative line flux seen in Abell 2390 into a luminosity,
and a mass, in the high and low density limits.  The total observed flux
would be 7.3$\times$10$^{{\rm -19}}$ erg cm$^{{\rm -2}}$ s$^{{\rm -1}}$, 
or a luminosity of 9.2$\times$10$^{{\rm 37}}$ erg
s$^{{\rm -1}}$ (about four orders of magnitude less than the luminosity of the optical
and H$_{{\rm 2}}$ lines).  In the limit where the density is greater than the critical
density (0.15 cm$^{{\rm -}{3}}$), the required gas mass is 1.2$\times$10$^{{\rm 10}}$ [N/H] M$_{\odot}$
and the size would need to be smaller than 11 kpc to achieve this density. 
This size is encompassed within the beam, which projects to a radius of
36 kpc.  The largest mass occurs when the emitting region uniformly fills
the beam, and this leads to a density below the critical density and a mass of 
6$\times$10$^{{\rm 10}}$ [N/H]$^{{\rm -}{\rm \frac{1}{2}}{}}$ (T/2$\times$10$^{{\rm 6}}$ K)$^{{\rm 1/4}}$ M$_{\odot}$.  The mass is confined to
a relatively narrow range that is orders of magnitude less than the total
gas mass in the system.  However, the cooling time for this gas is about 2
$\times$10$^{{\rm 6}}$ yr for a gas density of 0.1 cm$^{{\rm -3}}$, which would lead to an
enormous cooling rate in excess of 10$^{{\rm 3}}$ M$_{\odot}$ yr$^{{\rm -}{1}}$.  
For hot X-ray emitting gas to reach the temperature at which \ion{N}{7}
occurs, it would have to pass through the temperature range in which \ion{O}{7} is found
(10$^6$ K), yet the X-ray upper limits on the \ion{O}{7} emission line strength
restrict the cooling rate to $<$ 50 M$_{\odot}$ yr$^{-1}$ \citep{pete03}.
It might be possible to find a model in which the \ion{O}{7} is supressed while
the \ion{N}{7} is visible, but until such a model is found, the current theoretical
expectations would suggest that this emission feature in Abell 2390 is unlikely to be real.

The other feature of note is the absorption feature in 4C+39.25, which can be converted into a column
density for comparison with expectations.  In the optically thin limit, we
can use the usual linear relationship between the fractional equivalent
width and the column density to obtain N(\ion{N}{7}) = 1.6$\times$10$^{{\rm 16}}$ cm$^{{\rm -2}}$. 
However, this value does not take into account the excitation rate corrections
due to the interaction with the cosmic microwave background, as discussed by \citet{suny06}.
At z = 0.17, this reduces the cross section by a factor of 0.29.  Including this, the 
implied total column density is, for an ionization fraction of 0.5, N =
1.3$\times$10$^{{\rm 21}}$ (Z/Z$_{\odot}$)$^{{\rm -1}}$ cm$^{{\rm -}{2}}$.  If this were to represent a modest
overdensity filament, and the filament is typically 2 Mpc in diameter (\citealt{spring05}; 
or greater, if we are looking along a filament), the gas density
would be 2$\times$10$^{{\rm -4}}$ cm$^{{\rm -3}}$. 
While this seems feasible, it is larger than the columns that \citet{hell98} predict.  We have
used the tempertures, densities, and metallicities from their simulations to
determine the highest column densities in their simulations, covering a
redshift range of $\bigtriangleup$z = 0.3, which is comparable to the total path
length sampled by all strong continuum sources ($\bigtriangleup$z = 0.43).  Their
simulation implies a maximum column of N(\ion{N}{7}) = 1$\times$10$^{{\rm 15}}$ cm$^{{\rm -}{2}}$,
about 50 times lower than the column associated with this feature.  Their
metallicity is 0.08 Z$_{\odot}$ at the location of this absorption feature, and it is
hard to imagine that it could be more than a factor of a few higher, which would not
make their predicted column detectable.
A similar conclusion is reached by \citep{suny06}, who also considered the recent
models of \citep{cen06b}.
Since the model prediction is more than an order of magnitude lower than the \ion{N}{7} column 
density associated with this feature, it is unlikely to be a real absorption feature.

In the 40-50 GH range, the sources chosen for absorption studies are 
the brightest sources available, but for emission line studies, the 
best targets are at a redshift too low to be accessible with the 
current receiver.  For example, the two best galaxy clusters would 
be Abell 2597 and Abell 426 (the Perseus cluster), as both show 
strong evidence for cooling gas, at rates of 30-50 
M$_{{\rm \odot}{}}$ yr$^{-1}$ \citep{morr05, breg06}.
However, their redshifts place the NVII line at 49.0 GHz and 52.2 GHz.  
Observations at 52.2 GHz may never be feasible due to atmospheric 
extinction, but observations at 49.0 GHz may be possible with 
modifications to the GBT receiver.  Similarly, there are nearer 
galaxies with very active star formation (i.e., 10035+4852, 
10565+2488) and nearer X-ray bright elliptical galaxies (i.e., 
NGC 6414 and KIG 412), but this would demand that observations 
be made in the 50-51 GHz range, which is currently not possible.  
At these frequencies, the opacity decreases with altitude, so 
observing at the ALMA site has several advantages, although in 
the current plan, a 50 GHz receiver will not be available.

For purposes of detecting absorption by dilute gas in the million degree
range, it is valuable to compare the columns detectable using the
hyperfine \ion{N}{7} line and the X-ray lines of \ion{O}{7} and \ion{O}{8}.  At X-ray
energies, \ion{O}{7} absorption is the best line to detect and toward the
brightest sources it has been detected (at z = 0) for N(\ion{O}{7}) of 10$^{{\rm 16}}$ cm$^{{\rm -}{2}}$. 
These exposures typically require 300 ksec to reach this detection
threshold.  Our \ion{N}{7} observations are sensitive to a column
density about six times larger, but since nitrogen is six times less common than oxygen, the
equivalent total gas column needs to be correspondingly greater (about a factor of 30). 
However, our limits were obtained in less than an hour of integration time,
two orders of magnitude shorter than the X-ray observations.  For
the radio observations, if the rms decreases with the square root of the
integration time, a 30 hour observation would yield a
sensitivity for \ion{N}{7} that is within a factor of 5 of 
the sensitivity of \ion{O}{7} measurements in the X-rays.

One future goal is to determine whether the rms of a radio continuum
continues to decrease with longer integration times.  The current
observations have an rms that is near the theoretical value, but to achieve
it, we used a two-step procedure to flatten the continuum and remove the
standing wave structure.  It would be immensely helpful if the amplitude
of the standing waves could be reduced at the feedhorn and receiver
region.  The most immediate goal is to either confirm or refute the two
strongest features, and this can be accomplished by longer integrations
and by using a variety of different settings when observing the same
object (e.g., different central frequency, changing the focus by fractional
wave amounts).

\acknowledgments

We would like to thank the GBT staff for their considerable help, and in
particular to Frank Ghigo, Ron Madellena, and Jay Lockman.  JNB
would like to thank Mort Roberts, Dave Hogg, and Robert Brown for
their encouragement in pursuing this project.  Partial financial support
was provided by the NRAO, which is operated under contract for the
NSF by Associated Universities, Incorporated.

\clearpage
\begin{deluxetable}{llllllllllllllc}
\tabletypesize{\scriptsize}
\rotate
\tablecaption{\ion{N}{7} Emission and Absorption Observations}
\tablewidth{0pt}
\tablehead{
\colhead{No.} & \colhead{Name} & \colhead{Type}
 & \colhead{RA} & \colhead{DEC} & \colhead{z} & \colhead{t$_{exp}$} & \colhead{T$_a$} 
& \colhead{band} & \colhead{${\nu}_{center}$} & \colhead{rms\tablenotemark{b}} & \colhead{Bin Width} 
& \colhead{1$\sigma$\tablenotemark{b}}& \colhead{Comments}\\ 
\colhead{} & \colhead{} & \colhead{} 
&  \colhead{} & \colhead{} & \colhead{} & \colhead{(min)} & \colhead{(K)}
& \colhead{} & \colhead{GHz} & \colhead{(mK)} & \colhead{(km s$^{-1}$)} 
& \colhead{} & \colhead{} 
}
\startdata
1 & PKS0405$-$12  & blazar; Sy1.2 & 04:07:48.4 & $-$12:11:37 & 0.5726 & 32 & 0.31 & 5 & 45.5 & 10.3 & 26.4 & 0.52 & CIV abs at z = 0.167 \\
2 & PKS 1345+12 & Sy2 & 13:47:33.3 & +12:17:24 & 0.1217 & 32 & 0.47 & 7 & 46.9 & 6.10 & 25.6 & 0.30 & IR bright \\
 &  &  &   &  &  & 32 & 0.41 & 8 & 47.6 & 5.20 & 25.2 & 0.26 & \\  
3 & Abell 1835 & Galaxy Cluster & 14:01:02.0 & +02:51:32 & 0.2532 & 24 &  & 1 & 42.4 & 4.76 & 28.3 & 0.24 &  \\
 &  &  &  &  &  & 21 &  & 2 & 43.1 & 4.26 & 27.8 & 0.21 &   \\
4 & Abell 2390 & Galaxy Cluster & 21:53:34.6 & +17:40:11 & 0.2280 & 76 &  & 2 & 43.25 & 2.13 & 27.7 & 0.11 &   \\
5 & 3C273 &  QSO, blazar & 12:29:06.7 & +02:03:09 & 0.1583 & 80 & 16.1 & 5 & 45.4 & 0.34 & 26.4 & 0.017 & minus 3C279  \\
 &  &  &  &  &  & 80 & 14.5 & 6 & 46.1 & 0.37 & 26.0 & 0.018 & minus 3C279 \\
 &  &  &  &  &  & 32 & 15.8 & 7 & 46.9 & 0.80 & 25.6 & 0.040 & minus 3C279 \\
 &  &  &  &  &  & 32 & 12.8 & 8 & 47.6 & 1.00 & 25.2 & 0.050 & minus 3C279 \\
6 & 3C279 & QSO, blazar & 12:56:11.1 & $-$05:47:22 & 0.5362 & 34 & 6.4 & 1 & 42.4 & 0.82 & 28.3 & 0.041 &minus 3C345+4C39.25 \\
 &  &  &  &  &  & 34 & 6.1 & 2 & 43.1 & 1.04 & 27.8 & 0.052 & minus 3C345+4C39.25 \\
 &  &  &  &  &  & 32 & 13.2 & 3 & 43.9 & 0.85 & 27.3 & 0.042 & minus 3C345+4C39.25 \\
 &  &  &  &  &  & 32 & 11.7 & 4 & 44.6 & 1.02 & 26.9 & 0.051 & minus 3C345+4C39.25 \\
 &  &  &  &  &  & 32 & 11.6 & 5 & 45.4 & 0.34 & 26.4 & 0.017 & minus 3C273 \\
 &  &  &  &  &  & 32 & 10.5 & 6 & 46.1 & 0.37 & 26.0 & 0.018 & minus 3C273 \\
 &  &  &  &  &  & 44 & 3.5 & 7 & 46.9 & 0.80 & 25.6 & 0.040 & minus 3C273 \\
 &  &  &  &  &  & 44 & 2.9 & 8 & 47.6 & 1.00 & 25.2 & 0.050 & minus 3C273 \\
7 & 3C345 & QSO & 16:42:58.8 & +39:48:37 & 0.5928 & 32 & 3.8 & 1 & 42.4 & 0.89 & 28.3 & 0.044 & minus 3C279 \\
 &  &  &  &  &  & 32 & 3.5 & 2 & 43.1 & 1.02 & 27.8 & 0.051 & minus 3C279 \\
 &  &  &  &  &  & 30 & 1.9 & 3 & 43.9 & 1.09 & 27.3 & 0.054 & minus 3C279 \\
 &  &  &  &  &  & 30 & 1.7 & 4 & 44.6 & 1.29 & 26.9 & 0.064 & minus 3C279 \\
 &  &  &  &  &  & 32 & 3.9 & 5 & 45.4 & 0.75 & 26.4 & 0.038 & minus 3C279+3C273\\
 &  &  &  &  &  & 32 & 3.6 & 6 & 46.1 & 0.64 & 26.0 & 0.032 & minus 3C279+3C273 \\
 &  &  &  &  &  & 24 & 3.6 & 7 & 46.9 & 0.96 & 25.6 & 0.048 & minus 3C279+3C273 \\
 &  &  &  &  &  & 24 & 3.0 & 8 & 47.6 & 1.16 & 25.2 & 0.058 & minus 3C279+3C273 \\
\tablebreak
8 & 4C39.25 & QSO & 09:27:03.0 & 39:02:21 & 0.6953 & 32 & 6.4 & 1 & 42.4 & 0.75& 28.3 & 0.037 & minus 3C279 \\
 &  &  &  &  &  & 32 & 5.7 & 2 & 43.1 & 1.05 & 27.8 & 0.052 & minus 3C279 \\
 &  &  &  &  &  & 32 & 3.7 & 3 & 43.9 & 0.61 & 27.3 & 0.030 & minus 3C279 \\
 &  &  &  &  &  & 32 & 3.2 & 4 & 44.6 & 0.75 & 26.9 & 0.037 & minus 3C279 \\
 &  &  &  &  &  & 64 & 2.8 & 5 & 45.4 & 0.41 & 26.4 & 0.020 & minus 3C279+3C273 \\
 &  &  &  &  &  & 60 & 2.6 & 6 & 46.1 & 0.62 & 26.0 & 0.031 & minus 3C279+3C273 \\
 &  &  &  &  &  & 32 & 2.1 & 7 & 46.9 & 1.21 & 25.6 & 0.060 & minus 3C279+3C273 \\
 &  &  &  &  &  & 32 & 1.7 & 8 & 47.6 & 1.38 & 25.2 & 0.069 & minus 3C279+3C273 \\
\enddata
\tablenotetext{a}{Values for Abell 1835, and Abell 2390 are in mK; for the
rest, they are in units of 10$^{-3}$ of the normalized continuum.}
\tablenotetext{b}{Values for Abell 1835, and Abell 2390 are in K km s$^{-1}$
while uncertainties for other objects are equivalent widths, in km s$^{-1}$. }
\end{deluxetable}

\clearpage

\begin{figure}
\plotone{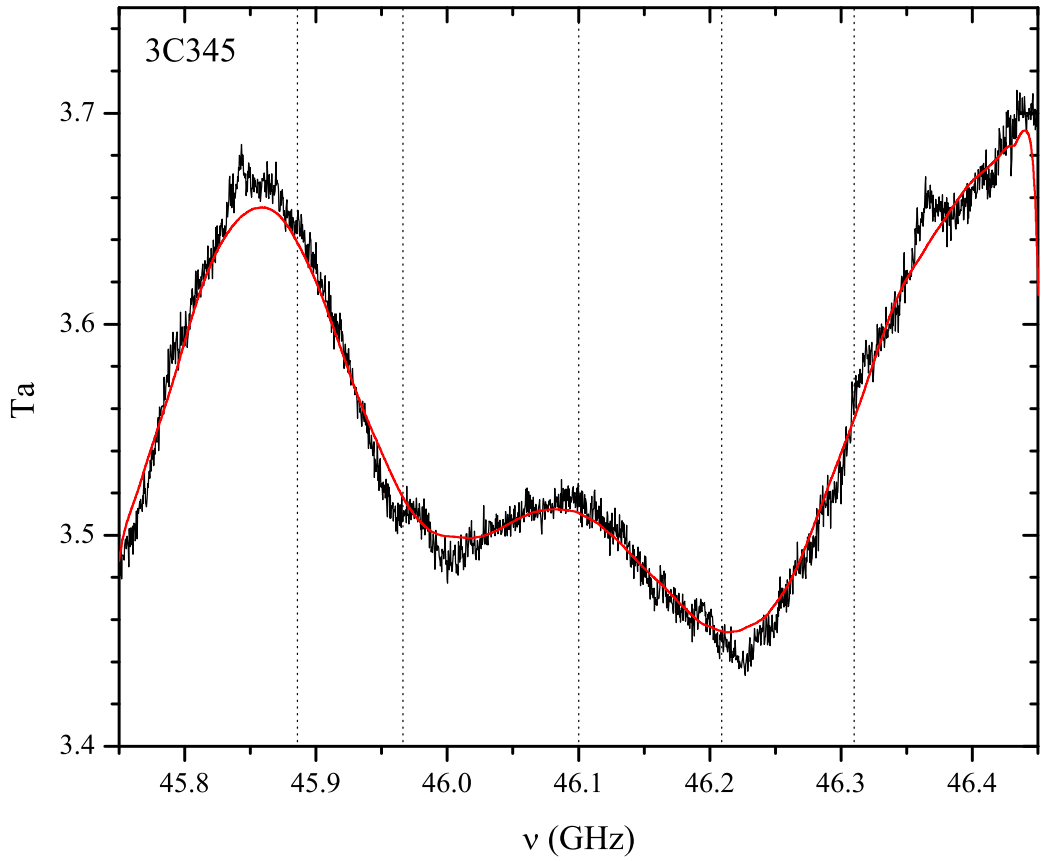}
\caption{The raw antenna temperature (K) from the summed scans for 3C 345 has
a 10\% variation across the band.  The data were smoothed by adjacent
averaging 200 bins, a width of about 0.08 GHz (smooth solid line).  To flatten the
spectrum, the first stage is to divide the data by the smoothed data.  The
dotted vertical lines show the locations of weak lines placed in the
spectrum during a test (maximum optical depth is 0.2\%).\label{fig1}}
\end{figure}

\begin{figure}
\plotone{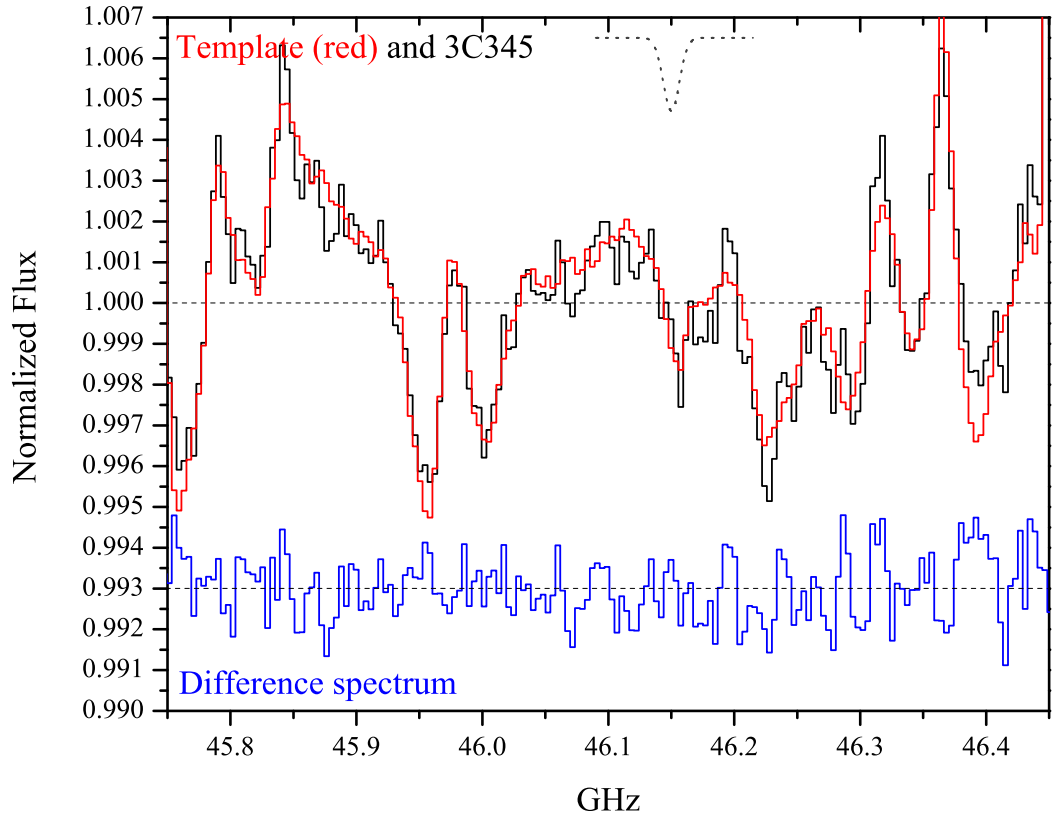}
\caption{
The ratio of the data to the smoothed data, in 10 channel bins (26 km s$^{{\rm -}{1}}$)
for 3C 345 and for the average of 3C 273 and 3C 279. 
The initial 10\% variations are reduced by an order of magnitude.  The
fluctuations about the unity line are subtracted from the normalized flux
of 3C 345 (the Difference spectrum), shown at the bottom of the figure
and shifted downward by 0.007.  This reduces the fluctuations by another
factor of five in this case, and the resulting rms is similar to the theoretical
limit.  The type of line that we are searching for is shown as a dotted line
in the top of the figure (90 km s$^{{\rm -}{1}}$ FWHM, central optical depth of 0.2\%).\label{fig2}}
\end{figure}

\begin{figure}
\plotone{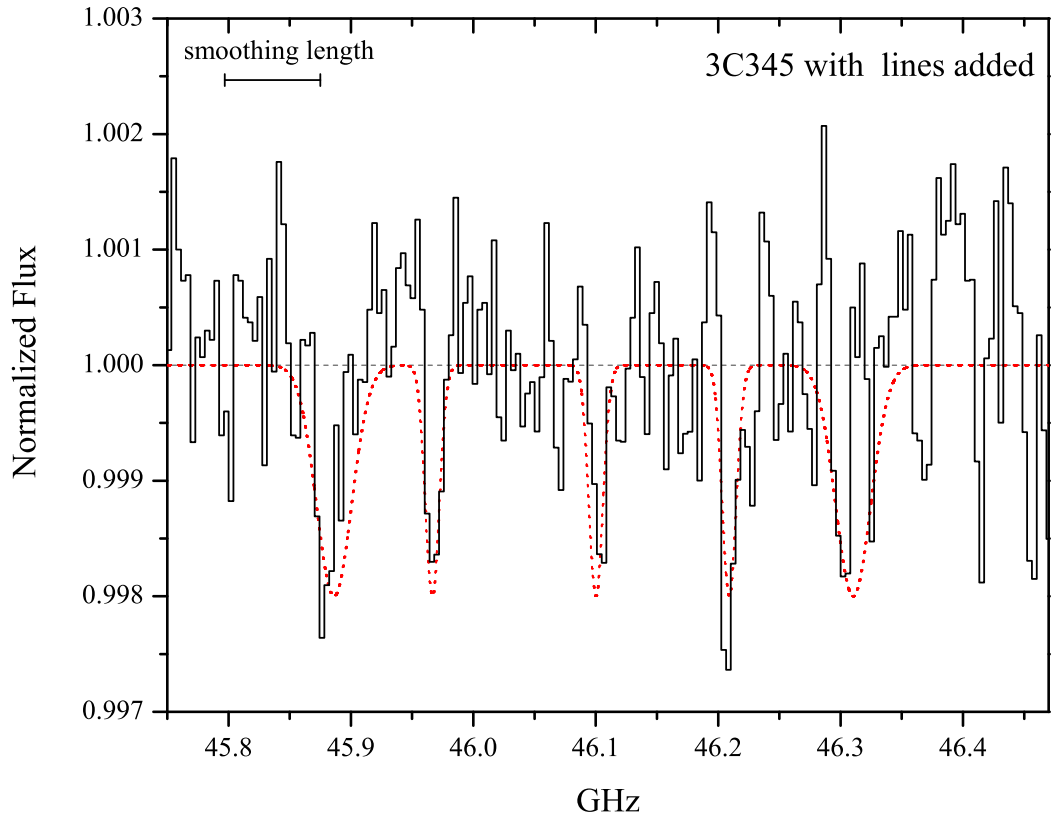}
\caption{
The same data set for 3C 345 as above, put processed with five weak
lines added to the spectrum of two different widths (dotted lines; FWHM
of 90 and 200 km s$^{{\rm -}{1}}$).  The five strongest absorption features are the
input lines, which would be recovered by our data reduction procedure. 
The smoothing length is shown in the upper left.}
\end{figure}

\begin{figure}
\plotone{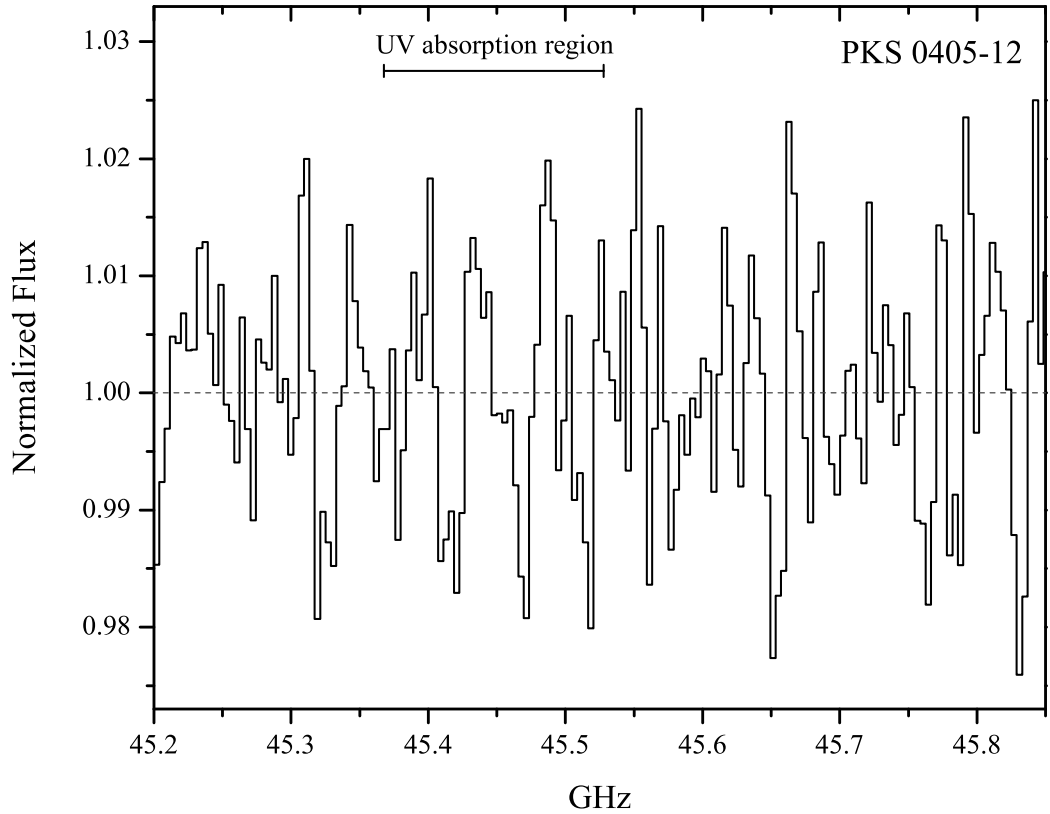}
\caption{
The normalized continuum of PKS 0405-12 does not show \ion{N}{7}
absorption at the location of the UV absorption lines, which are sensitive
to much smaller column densities.  The bar is 100 km s$^{{\rm -}{1}}$ in length,
centered at the UV absorption line redshift.}
\end{figure}

\begin{figure}
\plotone{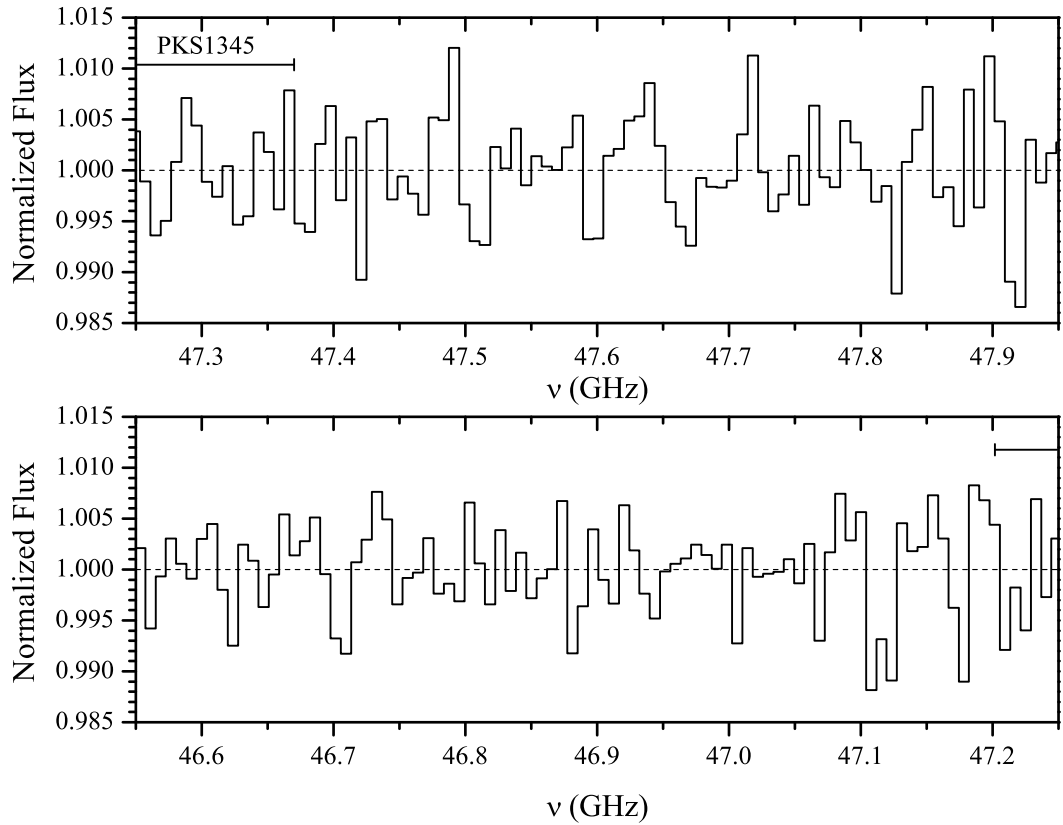}
\caption{
The normalized continuum from the IR-bright double nucleus souce PKS
1345+12 shows no emission or absorption from \ion{N}{7}.  The horizontal
bar is 1000 km s$^{{\rm -}{1}}$ in length and is centered on the redshift of the object.}
\end{figure}

\begin{figure}
\plotone{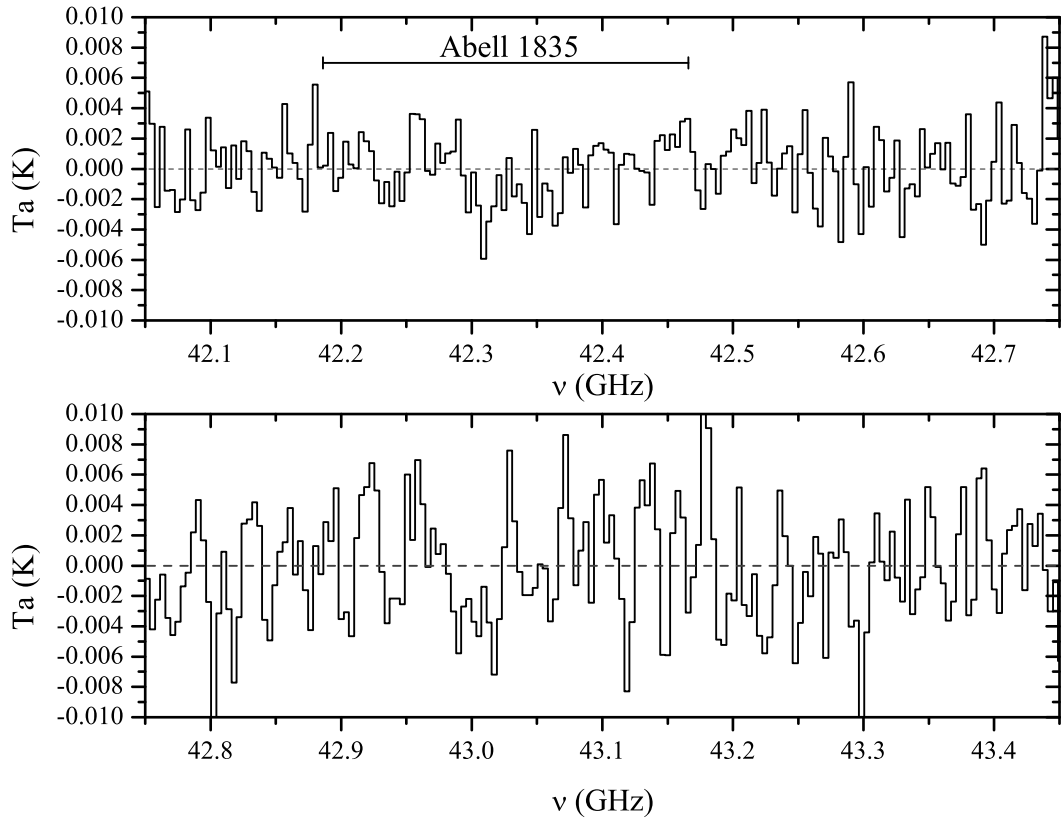}
\caption{
The observations from the X-ray bright galaxy cluster Abell 1835 shows
no \ion{N}{7} emission.  A region 2000 km s$^{{\rm -}{1}}$ wide is denoted with a
horizontal bar centered at the cluster redshift.}
\end{figure}

\begin{figure}
\plotone{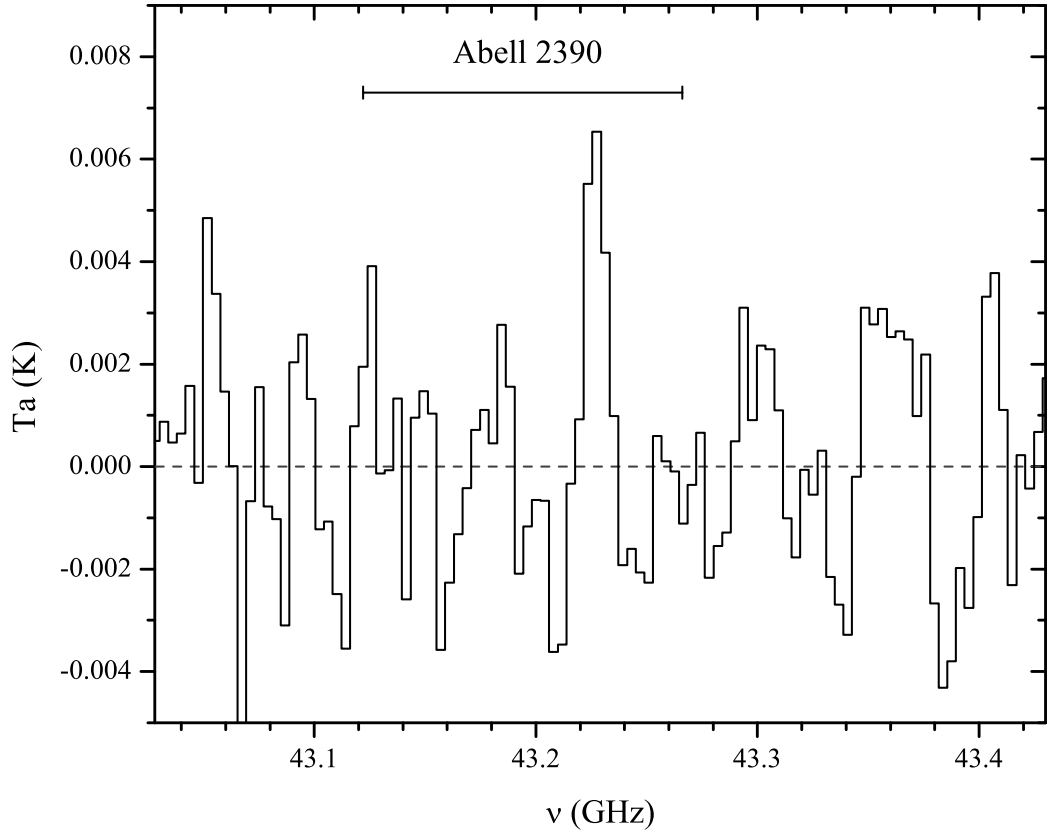}
\caption{
The observations of Abell 2390 show a 3.7$\sigma$ feature that is consistent
with the redshift of the cluster, where the horizontal bar is 1000 km s$^{{\rm -}{1}}$ in
length.}
\end{figure}

\begin{figure}
\plotone{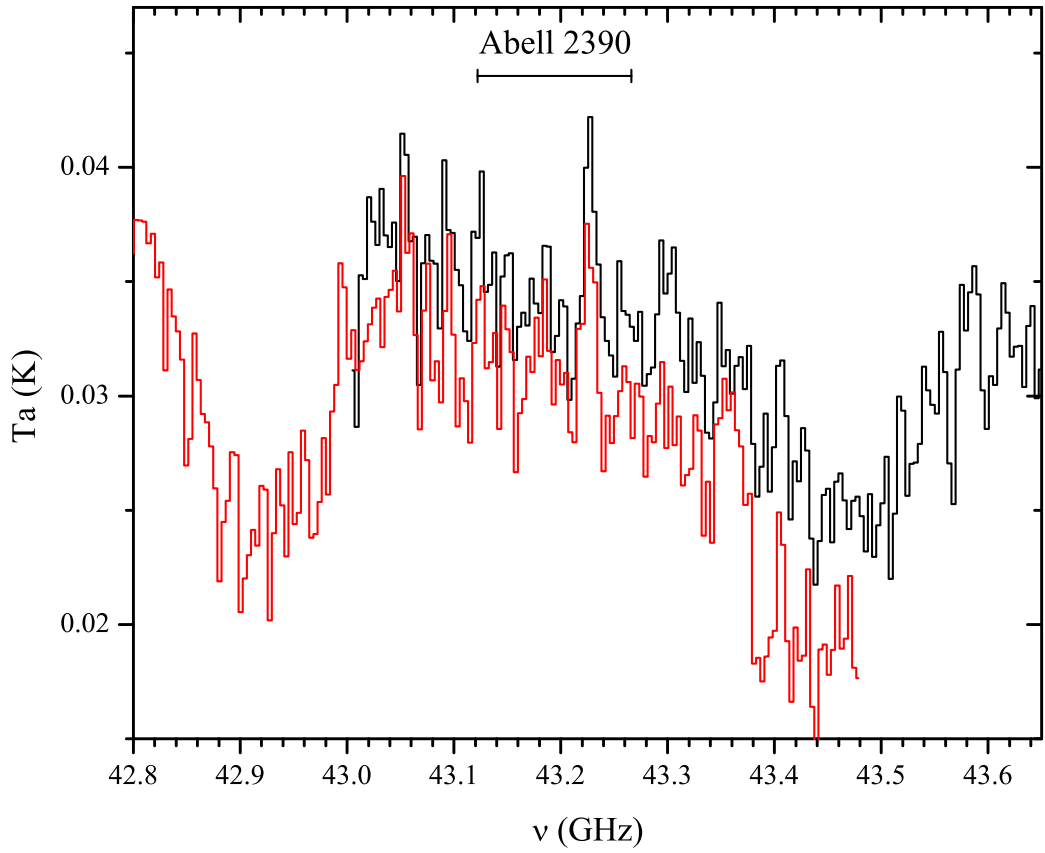}
\caption{
The two overlapping spectroscopic observations of Abell 2390, before
baseline removal.  The strongest emission feature is present in both
observations; the horizontal bar is 1000 km s$^{{\rm -}{1}}$ wide. }
\end{figure}

\begin{figure}
\plotone{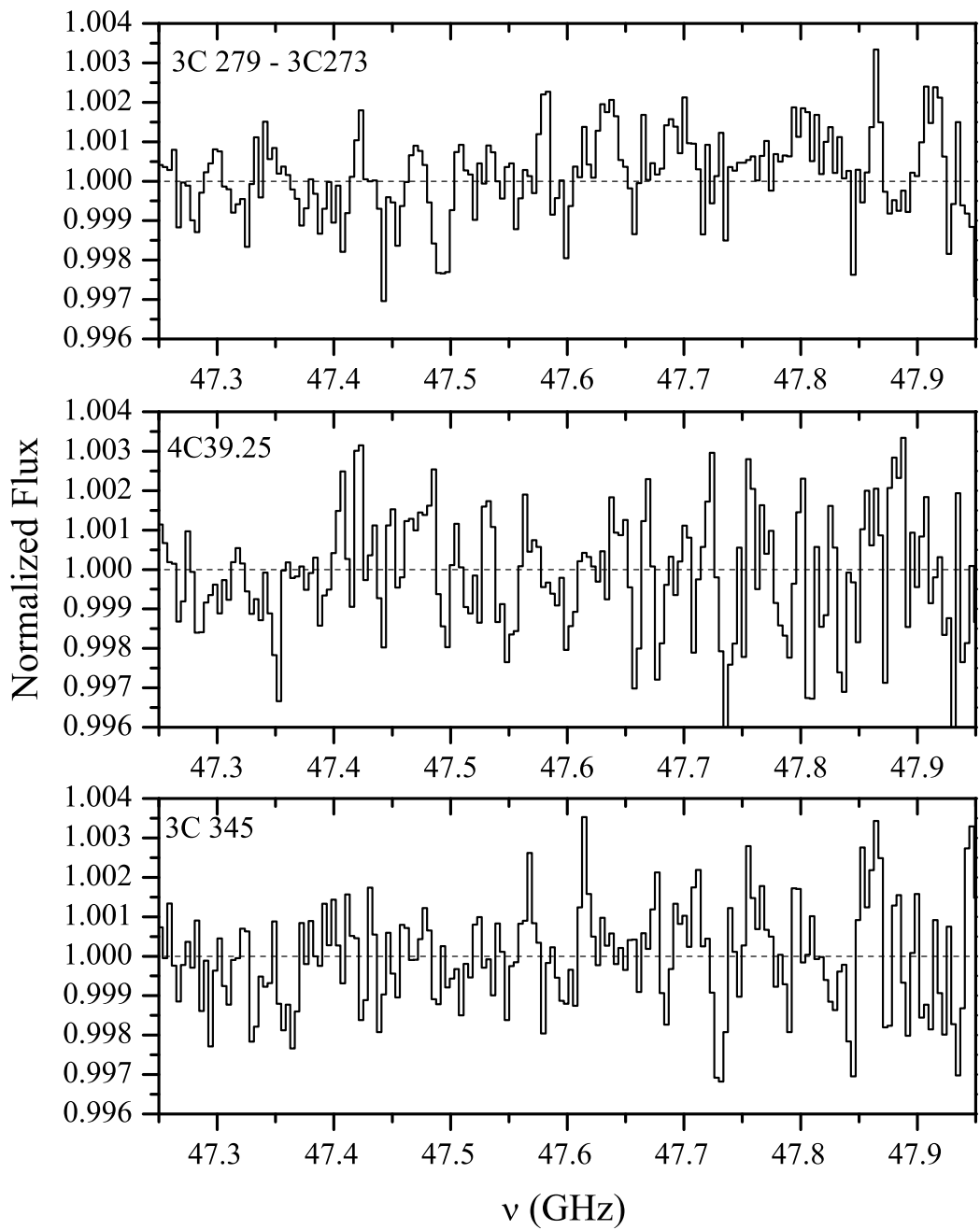}
\caption{
The flattened continuum sources in the 47.25 - 47.95 GHz region (band 8; 0.107
$<$ z $<$ 0.124 for the \ion{N}{7} line).  In the second stage of baseline flattening,
3C 273 and 3C279 were differenced, while for 3C+39.25 and 3C 345,
small baseline residuals were removed by using the average of 3C 273 and
3C 279.  The total vertical range is $\pm$0.4\%.}
\end{figure}

\begin{figure}
\plotone{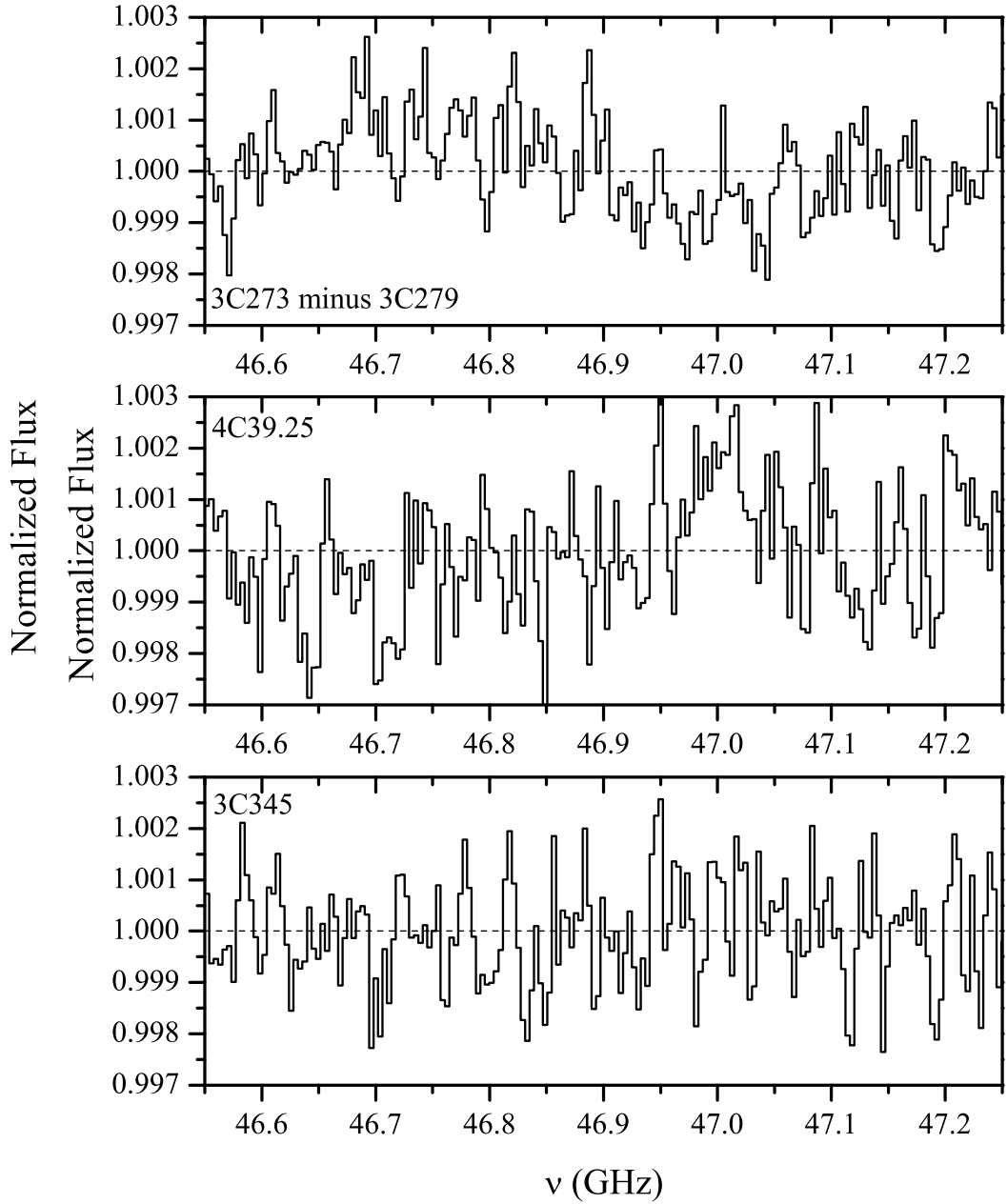}
\caption{
As above, but for the 46.55 - 47.25 GHz range (band 7; 0.124 $<$ z $<$ 0.141).  The
total vertical range is $\pm$0.3\%.}
\end{figure}

\begin{figure}
\plotone{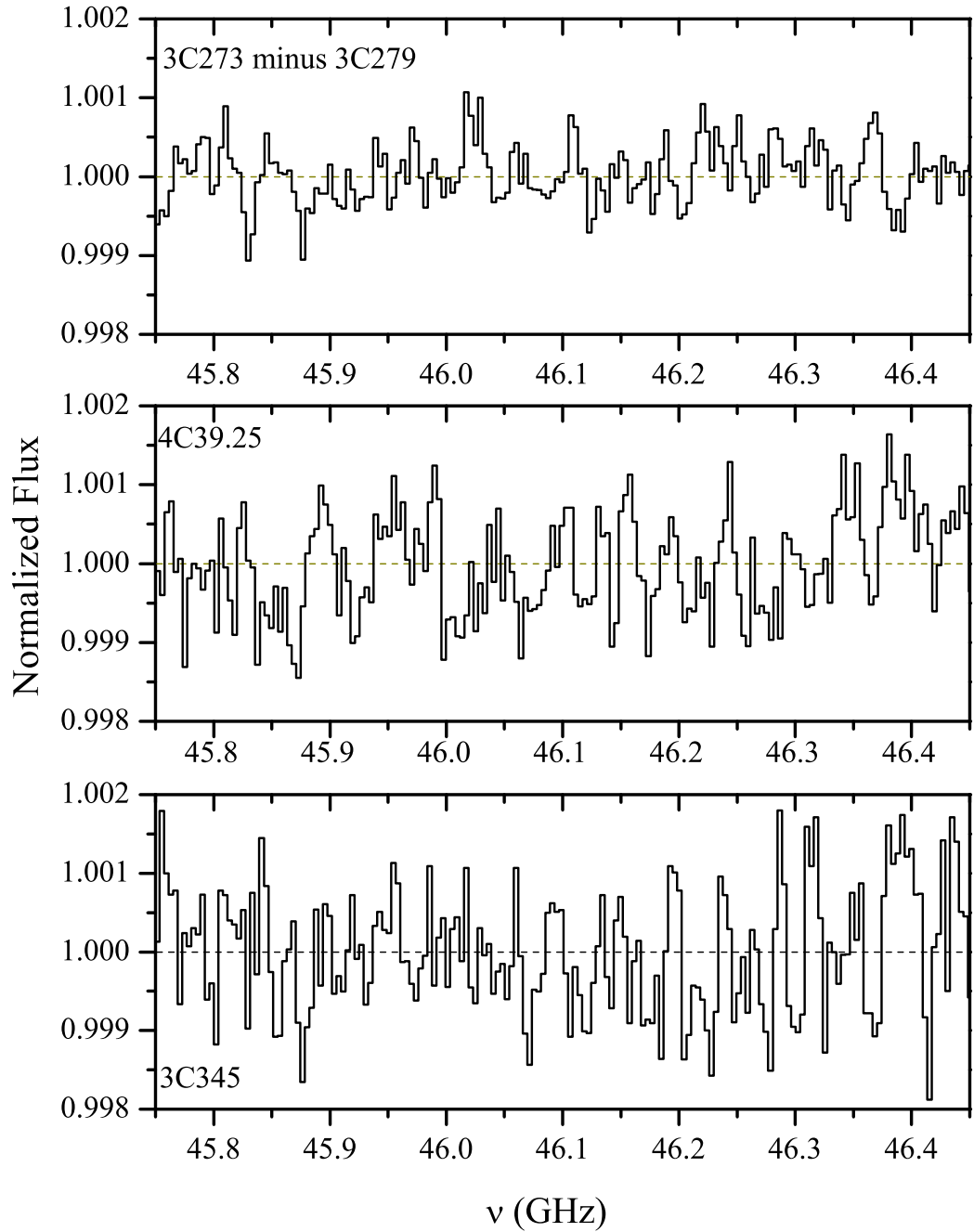}
\caption{
As above, but for the 45.75 - 46.45 GHz range (band 6; 0.143 $<$ z $<$ 0.161).  The
total vertical range is $\pm$0.2\%.}
\end{figure}

\begin{figure}
\plotone{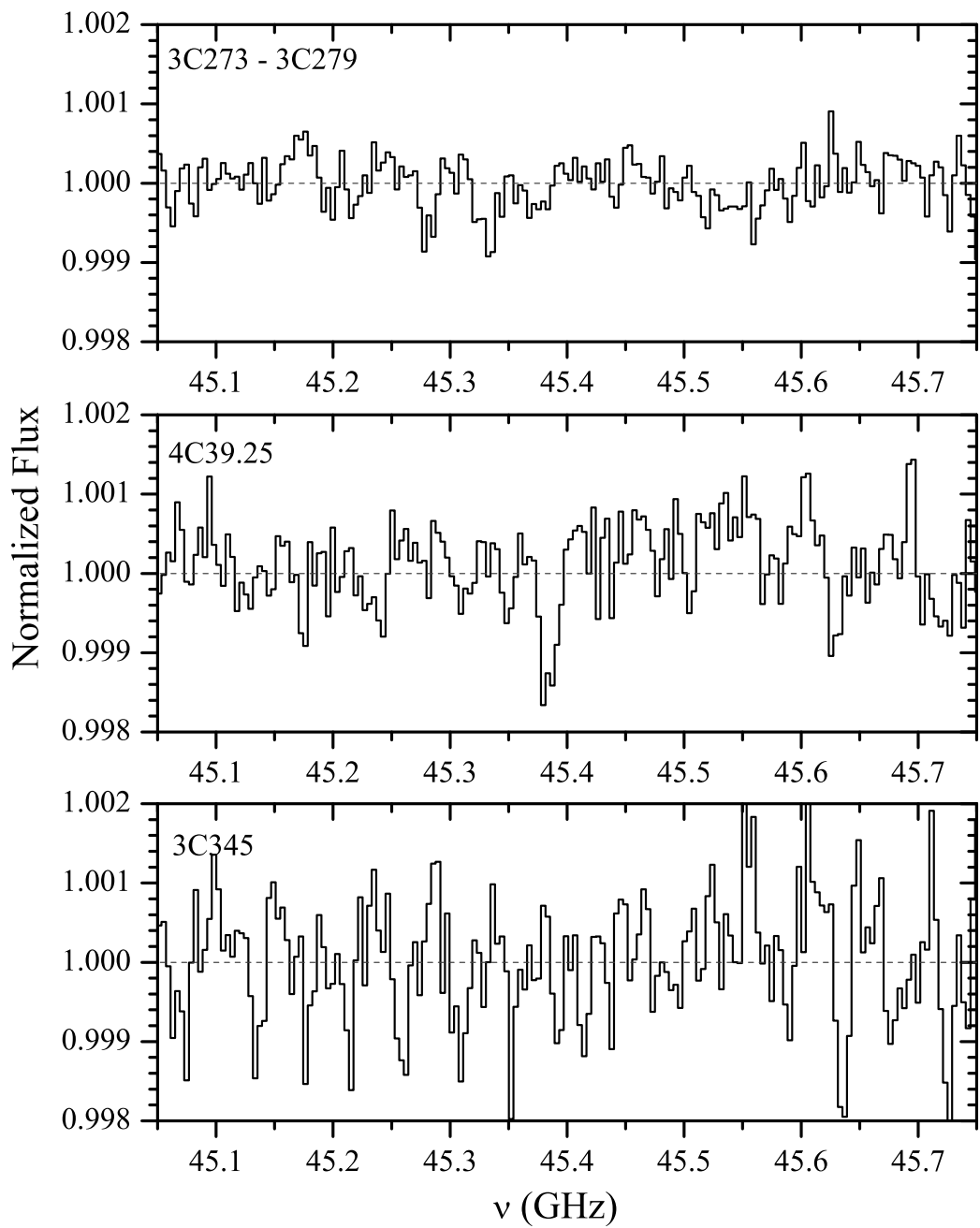}
\caption{
As above, but for the 45.05 - 45.75 GHz range (band 5; 0.161 $<$ z $<$ 0.179).  The
total vertical range is $\pm$0.2\%.  The strongest absorption feature in any spectrum
occurs in 4C+39.25 at 45.38 GHz, a 3.8$\sigma$ feature.}
\end{figure}

\begin{figure}
\plotone{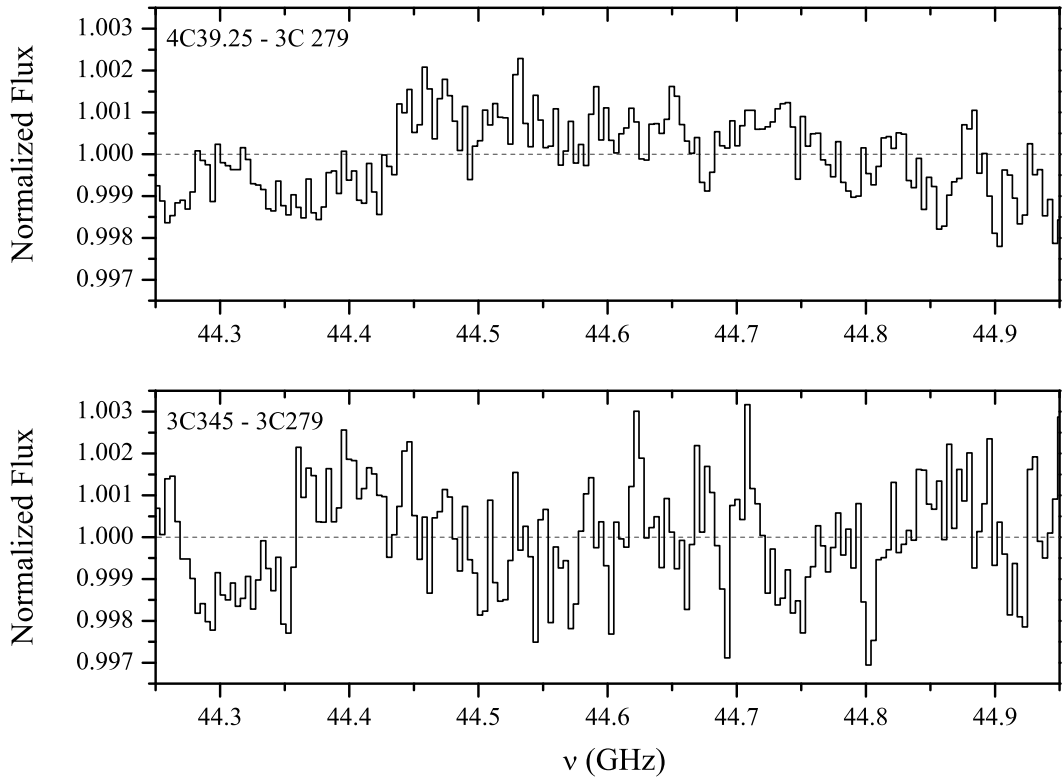}
\caption{
As above, but for the 44.25 - 44.95 GHz range (band 4; 0.181 $<$ z $<$ 0.200).  The
source 3C 273 was not observed below 45 GHz, so the continuum of the
strong source 3C 279 was used to remove small baseline residuals from
4C+39.25 and 3C 345.  The total vertical range is $\pm$0.35\%.}
\end{figure}

\begin{figure}
\plotone{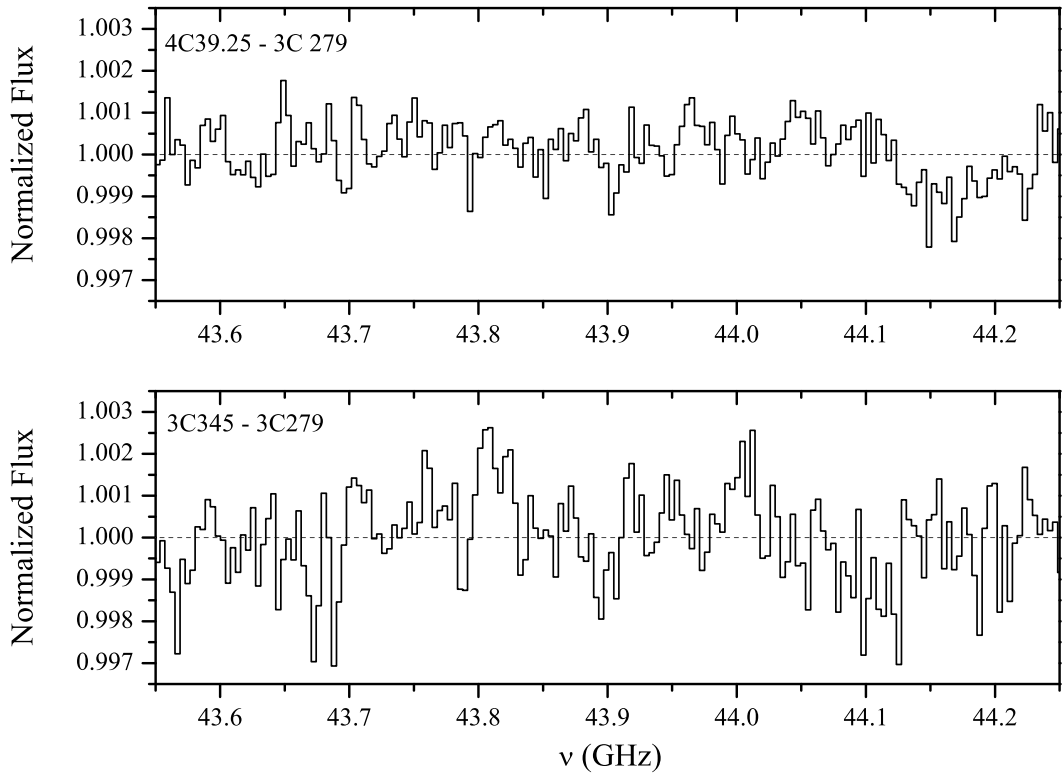}
\caption{
As above, but for the 43.55 - 44.25 GHz range (band 3; 0.200 $<$ z $<$ 0.219).}
\end{figure}

\begin{figure}
\plotone{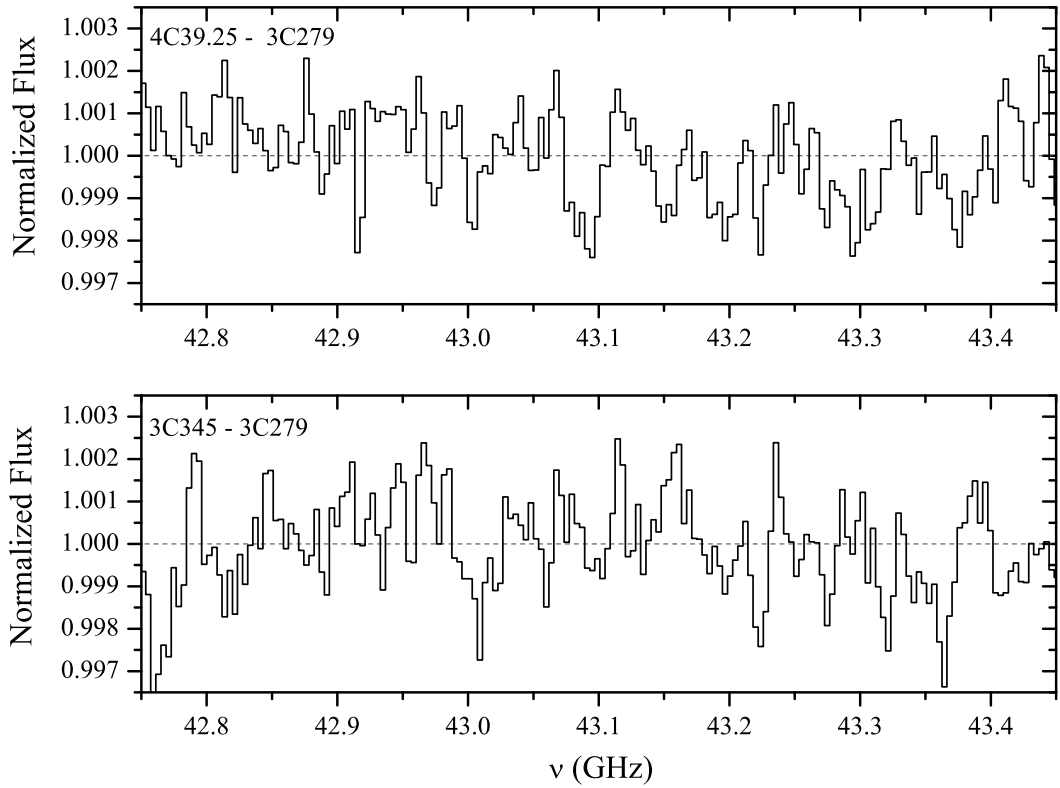}
\caption{
As above, but for the 42.75 - 43.45 GHz range (band 2; 0.222 $<$ z $<$ 0.242).  The
feature in the 4C+39.25 spectrum near 43.1 GHz is a 2.7$\sigma$ feature and
such features are expected for the amount of data obtained and the number of
independent spectral resolution elements.}
\end{figure}

\begin{figure}
\plotone{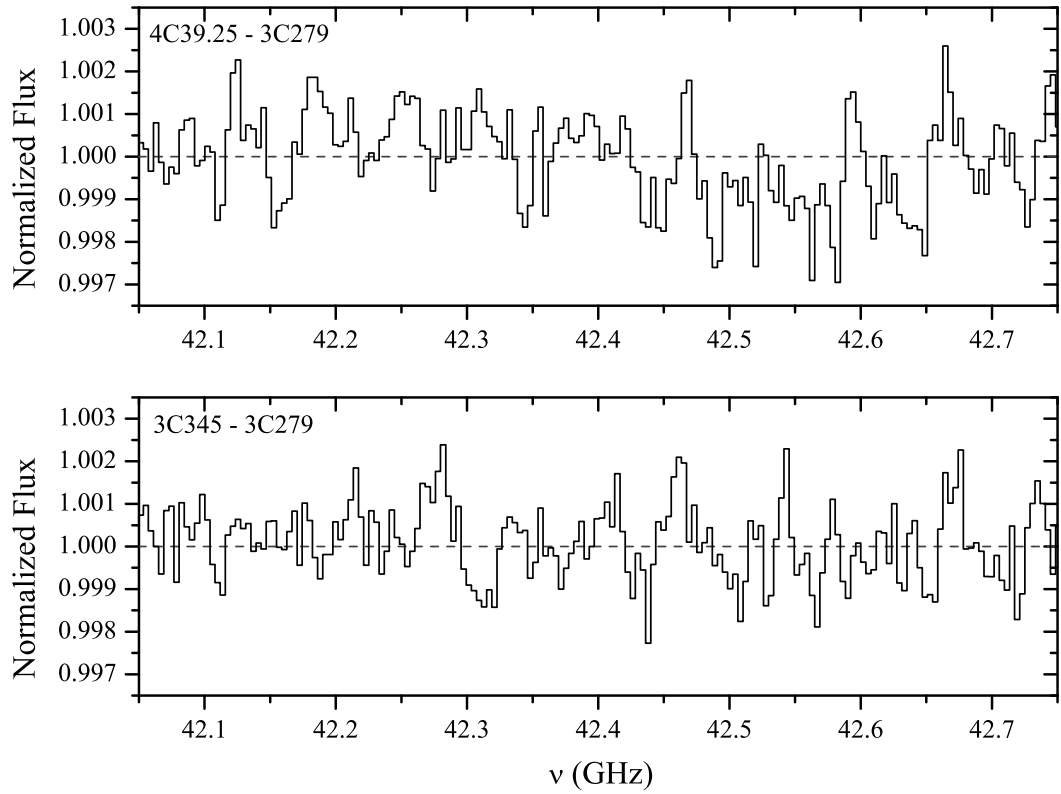}
\caption{
As above, but for the 42.05 - 42.75 GHz range (band 1; 0.242 $<$ z $<$ 0.263).}
\end{figure}


\begin{thebibliography}{}

\bibitem[Allen et al.(2001a)]{allen01a} Allen, S.~W., Ettori, S., 
\& Fabian, A.~C.\ 2001, \mnras, 324, 877 

\bibitem[Allen et al.(2001b)]{allen01b} Allen, S.~W., Schmidt, 
R.~W., \& Fabian, A.~C.\ 2001, \mnras, 328, L37 

\bibitem[Bregman et al.(2006)]{breg06} Bregman, J.~N., Fabian, 
A.~C., Miller, E.~D., \& Irwin, J.~A.\ 2006, \apj, 642, 746 

\bibitem[Cen \& Fang(2006)]{cen06b} Cen, R., \& Fang, T.\ 
2006, \apj, 650, 573 

\bibitem[D'Cruz et al.(1998)]{dcru98} D'Cruz, N.~L., Sarazin, 
C.~L., \& Dubau, J.\ 1998, \apj, 501, 414 

\bibitem[Edge et al.(2002)]{edge02} Edge, A.~C., Wilman, 
R.~J., Johnstone, R.~M., Crawford, C.~S., Fabian, A.~C., \& Allen, S.~W.\ 
2002, \mnras, 337, 49


\bibitem[Garwood et al.(2005)]{garw05} Garwood, R., Braatz, 
J., Marganian, P., \& Radziwill, N.\ 2005, American Astronomical Society 
Meeting Abstracts, 207, \#29.08 

\bibitem[Goddard \& Ferland(2003)]{godd03} Goddard, W.~E., \& 
Ferland, G.~J.\ 2003, \pasp, 115, 647 

\bibitem[Hellsten et al.(1998)]{hell98} Hellsten, U., Gnedin, 
N.~Y., \& Miralda-Escud{\'e}, J.\ 1998, \apj, 509, 56 

\bibitem[Liang et al.(1997)]{lian97} Liang, H., Dickey, J.~M., 
Moorey, G., \& Ekers, R.~D.\ 1997, \aap, 326, 108 

\bibitem[Morris \& Fabian(2005)]{morr05} Morris, R.~G., \& 
Fabian, A.~C.\ 2005, \mnras, 358, 585 

\bibitem[Peterson et al.(2003)]{pete03} Peterson, J.~R., Kahn, 
S.~M., Paerels, F.~B.~S., Kaastra, J.~S., Tamura, T., Bleeker, J.~A.~M., 
Ferrigno, C., \& Jernigan, J.~G.\ 2003, \apj, 590, 207 

\bibitem[Rasmussen et al.(2006)]{rasm06} Rasmussen, A.~P., 
Kahn, S.~M., Paerels, F., Willem den Herder, J., Kaastra, J., \& de Vries, 
C.\ 2006, ArXiv Astrophysics e-prints, arXiv:astro-ph/0604515 

\bibitem[Shabaev et al.(1995)]{shab95} Shabaev, V.~M., 
Shabaev, M.~B., \& Tupitsyn, I.I.\ 1995, \pra, 52, 3686 

\bibitem[Spinrad et al.(1993)]{spin93} Spinrad, H., et al.\ 
1993, \aj, 106, 1 

\bibitem[Springel et al.(2005)]{spring05} Springel, V., et al.\ 
2005, \nat, 435, 629

\bibitem[Sunyaev \& Churazov(1984)]{suny84} Sunyaev, R.~A., 
\& Churazov, E.~M.\ 1984, Soviet Astronomy Letters, 10, 201 

\bibitem[Sunyaev \& Docenko(2006)]{suny06} Sunyaev, R.~A., \& 
Docenko, D.\ 2006, ArXiv Astrophysics e-prints, arXiv:astro-ph/0608256 

\bibitem[Xiang et al.(2002)]{xian02} Xiang, L., Stanghellini, 
C., Dallacasa, D., \& Haiyan, Z.\ 2002, \aap, 385, 768 

\end{thebibliography}
\end{document}